# FERMI STATISTICS METHOD APPLIED TO MODEL MACROSCOPIC DEMOGRAPHIC DATA

**Giuseppe Alberti**
Independent Researcher

**email: giuseppe.alberti@squ-systems.eu**
**ORCID: 0000-0002-3041-8016**

**Abstract**

*The study begins by considering an abstract object (cellular automaton) able of moving -by arbitrary decision- between two given fixed positions. That is, at each clock step, it can change position or remain stationary in its current position. This object, which we call an Arbitrary Oscillator (ArbO), cannot evolve indefinitely since it may encounter 'end-of-life' events, which are also random. If we place quantitative limits on the number of arbitrary events and impose that the life cycle of ArbO must end in any case, we can use Fermi statistics to find the most probable distribution of fatal events along the possible sequences of choices. This distribution is represented by a recursive function that can be calculated for each total number of possible 'life/death' choices, which we will call Total Cases (TC). By means of a time-scale adjustment, we have associated the distribution curves of ArbO 'fatal' events with the demographic mortality curves (dx and qx data) of populations in the case of Italy. To better study the properties of the statistical function thus found, we attempted a continuous transposition of the recursive equation, seeking solutions to the differential equation linkable with it. With a continuous analytical expression, the characteristics of this statistical distribution can be studied more effectively. Similarities and differences with demographic mortality curves have been highlighted, attempting to explain the latter as overlaps of curves with different TC parameters. Implications with life span and more general life cycle concepts are outlined. A correlation with a more recent study using a multi-omics approach is also pointed out. Key Words: Cellular Automata, Fermi Statistics, Logistic Distribution, Demographic Mortality, Lifespan*

## 1. Introduction

A great deal of study has been done on the topic of demographic mortality particularly at advanced ages. Accurate estimates of life span prediction have obvious social and economic motivations. Many references were made in these studies to concepts coming from various scientific disciplines, such as references to reliability theory, computer theories, biological models, entropy theory, and many others. Ref. [1-6]. The present study aims to bring a point of view that -in the author's opinion- has some new aspects. The original features of the work are the use of statistical mechanics methods and the presence of only one parameter in the model equations. The basic finding comes from the study of a cellular automaton model, when a recurring equation is found as follows:

$$m(\mathrm{TC},\ r) = \left(2^r - \sum_{t=1}^{r-1} 2^{r-t}\, m(\mathrm{TC},\ t)\right) \Big/ (1 + 2^{-r}(\mathrm{TC} - 2)); \; m(\mathrm{TC},\ 1) = 0;\; r \geq 2; \qquad (1)$$

where the TC ("Total Cases") symbol is a parameter, and *r* represents the recursion step. If we associate the above (1) function with some mortality statistical data vs time intervals (addressed by the *r* step) and if we assume the total cases sample as TC, we found the Fig.1 with TC=100 000 . Note that the TC value corresponds to the area under both curves. The "*m*" curve is generated by the (1) function, while the other curve corresponds to the mortality table data available at the ISTAT web site for e.g. Italy 2019 when the r steps correspond to a five years interval count and the vertical axes represent the deaths in the r interval for a total sample of 100 000 people Ref. [7]. The two curves show a quite evident similarity, having a peak close feature and in particular a similar descending slope. The major difference is in the leading part where the curves show different slopes. It must be noted that the "*m*" curve is not a fitting attempt on the ISTAT curve but it is just an independent deterministic computation of the (1) function when TC=100 000 and the r step runs the positive integer range.

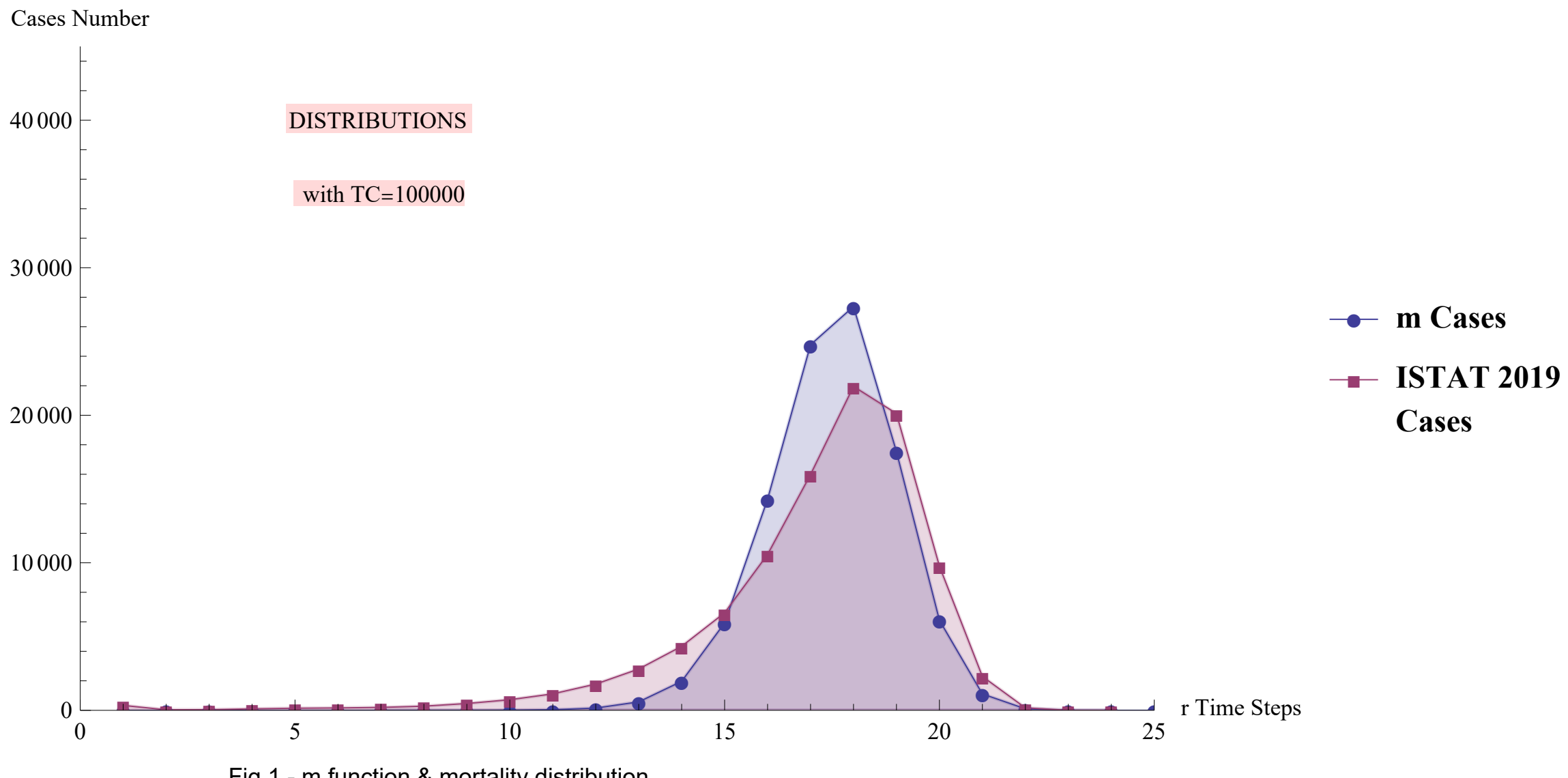

Fig.1 - m function & mortality distribution

The next section will introduce the cellular automaton at the origin of the above recursive equation. The third section will discuss the mathematics of the object, and subsequent sections will present a Fermi's statistics method application to the model leading to the proposed recursive algorithm. A comparison with demographic curve data will then be provided. To further explore the features and capabilities of the model, an additional section will introduce continuous equations instead of recursive finite difference equations. Finally, the features of the model will be discussed leading to possible explanations of the shape of the demographic curves.

# 2. The arbitrary oscillator

To build the equation (1) we need to introduce a particular cellular automaton. Consider the Fig. 2

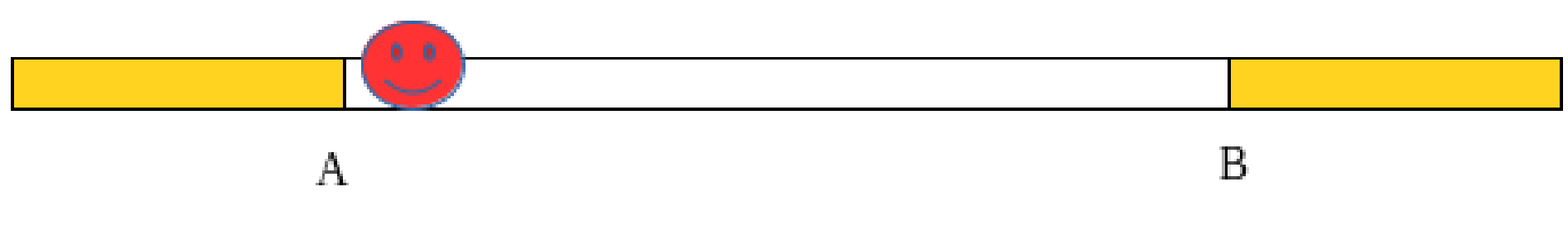

Fig.2 – Arbitrary oscillator

In the figure a "bug" can move on a horizontal x axis exclusively between two stable positions points: A and B. The decision to move (or to stay) is free for the "bug", therefore not predictable. In the boundary positions the "insect" encounters the outside world that can determine a "safe" condition (food, reproduction,...) or a "deadly" condition (predators, danger, disease,...). These occurrences are also not predictable. At each step of time the "bug" decides what to do: stay/move (say "0" for stay and "1" for move). A "path-space" for this Arbitrary Oscillator (ArbO) can be envisaged. It includes a space and time coordinate and a "d" path index, collecting all the possible decisions sequence. Fig. 3 shows the scheme. The figure depicts a sequence of four time steps with binary decisions labeled "0" or "1" and amounting in 16 total possible paths (or “orbits”). The sequence start at a common initial spatial position. Note, at the sides of the figure, the “extreme” decisions possibilities : a ) {1,1,1,1} -> ever change or b) {0,0,0,0}-> stay still.

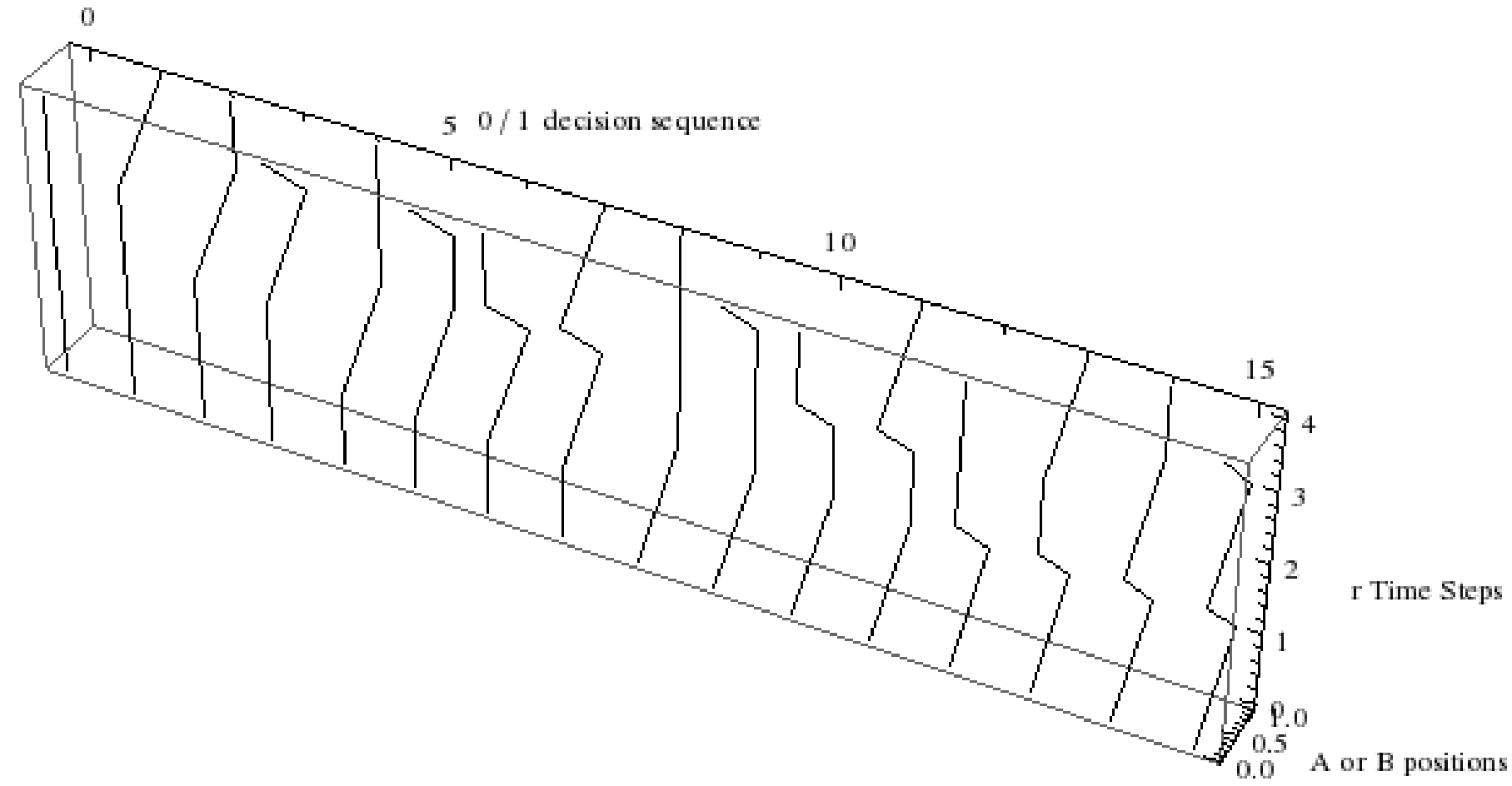

Fig.3 - path-space of the Arbitrary Oscillator for 4 steps of time decisions

The same situation is shown in Fig. 4 where the spatial coordinate is omitted. At each decision step, the rectangular “cell” splits into two subsequent underlying cells. By convention, the left new cell can represent a “no move” or “0” decision, while the right new cell results from a “move” or “1” decision. The total final possibilities are 16 different sequences classified by the “d” index shown at the bottom of the last group of cells. The d index label can follow any coding preferred rule. The cycle starts at an initial zero step, common to all subsequent paths, where the next decision is drawn, resulting in the two different cells of the first step and so on.

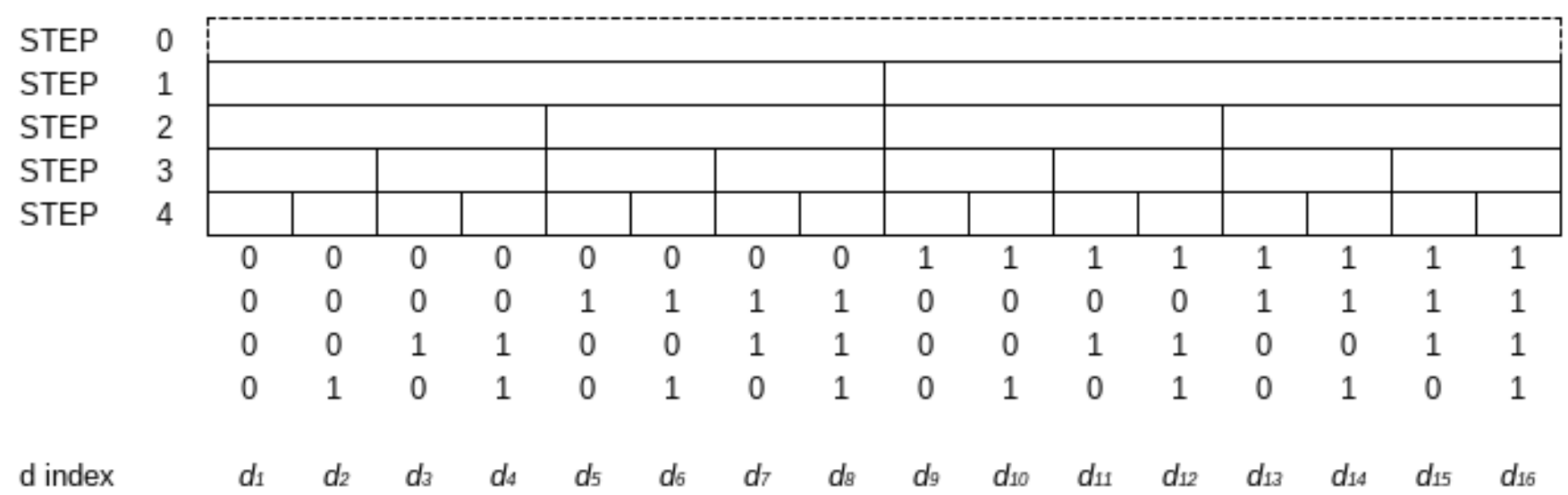

Fig.4 - Possible choices sequences

In the following Fig. 5, we introduce -for our ArbO- the risk of a End of Life (EoL) event. In this case, the crossed cells mean the occurring of a EoL event. Of course in this case no subsequent cells can be expected and the final “d” sequence is truncated. The EoL events are randomly distributed and can arise at any available decision cell, one per cell, on one or many cells in the level.

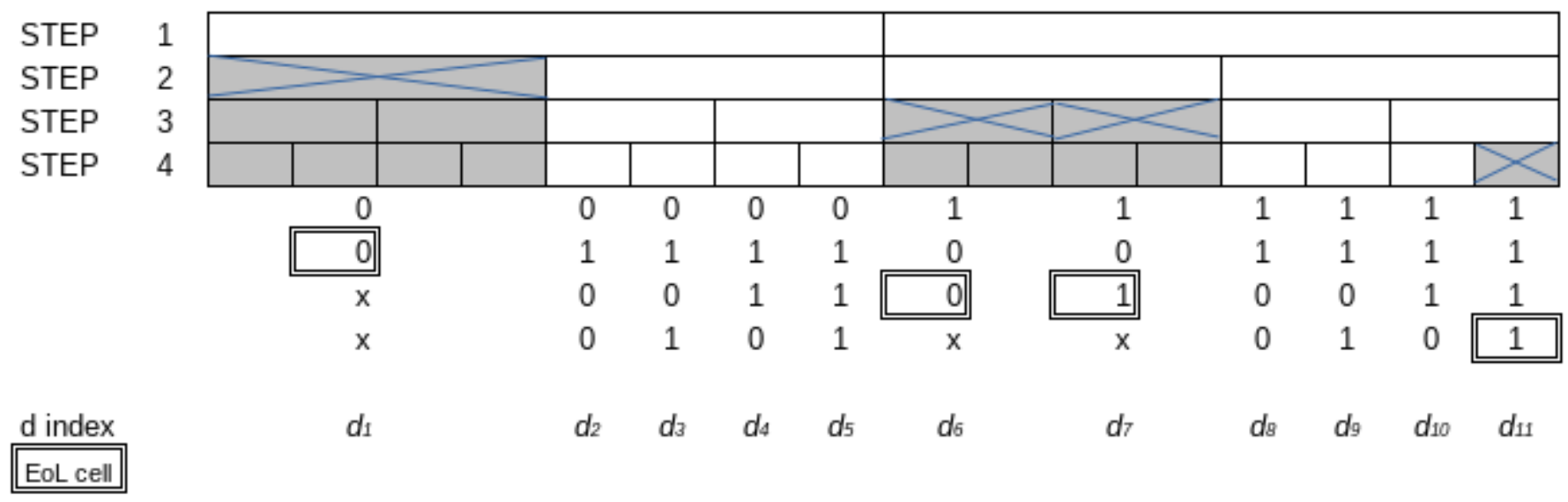

Fig. 5 - EoL events

With the EoL event possibility, an end of the full “life cycle” of the ArbO can be considered as per Fig. 6. Here the four step case is again presented. In the figure all the “path-space” is covered by sequences terminating with and EoL event. Note that, for the four step case in this example, we find a five different path history.

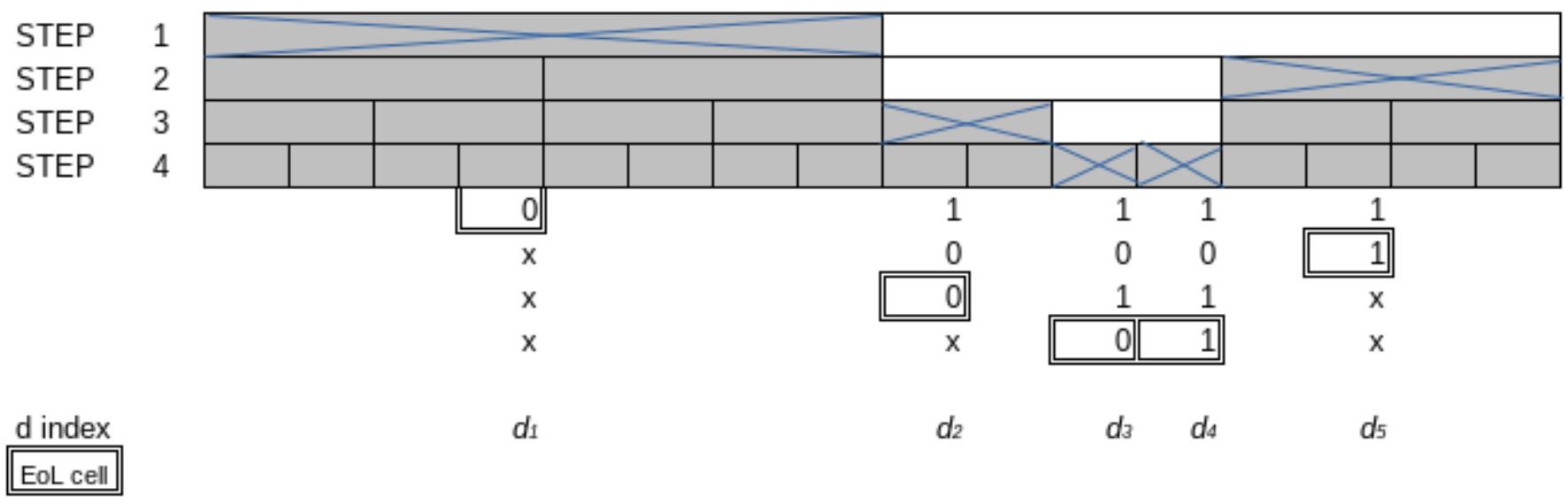

Fig.6 - A fully ended life cycle of the arbitrary oscillator

# 3. The arbitrary oscillator equations

The above processes can be mathematically described assuming that:

*(i) - The EoL events cover only one cell per event. Their occurrence is not predictable, in number and position, for any time step level and over the available level cells.*

*(ii) - The number of possible ArbO decision sequence paths is limited to a max Total Cases (TC).*

The *(i)* statement formalize the ArbO behavior described in the previous Section 2. The *(ii)* statement means simply that the step sequence cannot be “perpetual” i.e. unlimited. Consider indeed that after 116 steps (without EoL events) the number of possible sequences in the path space becomes about $10^{35}$. In this case -suppose be L a fixed length in the d space dimension- the mean distance between the paths will be $L*10^{-35}$. With *(i)* and *(ii)* assumptions, we can then define the following relations and conditions:

$$m_r + v_r = 2\, v_{r-1} \tag{2}$$

$$\begin{gathered} 0 \le m_r,\ v_r \le 2^r \\ m_r,\ v_r \text{ integers} \\ v_0 = 1;\ r \ge 1 \end{gathered} \tag{3}$$

We define $m_r$ as the number of EoL events at the step $r$ and $v_r$ the remaining “safe” decision cells at the same step level. Referring e.g. to Fig. 5 at step 2 level we have $m_2 = 1$ and $v_2 = 3$. The factor of 2, in the Eq. (2), comes from the “binary” decision rule.

Solving the recursion equation (2) vs the $m$ variables, we find:

$$v_r = 2^r - \sum_{t=1}^{r} 2^{r-t}\, m_t \tag{4}$$

From the statement (*ii*) it follows that any sequence of decisions must end with an EoL event. The total number of these terminated paths must be TC . Then:

$$\sum_{r=1}^{\text{Rmax}} m_r = \text{TC} \tag{5}$$

It is easy to demonstrate that:

$$\text{Rmax} = \text{TC} - 1 \tag{6}$$

The presence of the $v_r$ variables in the above relations can be eliminated if we consider the “final” level of the life cycle of our ArbO: there, it must be $v_{\text{Rmax}} = 0$ . Then factoring the term $2^r$ and renaming $t$ index as $r$ index the eq. (4) leads to the (7) equations, with equivalent expressions :

$$\begin{gathered} \sum_{r=1}^{\text{Rmax}} 2^{-r}\, m_r = 1 \\ \sum_{r=1}^{\text{Rmax}} 2^{\text{Rmax}-r}\, m_r = 2^{\text{Rmax}} \end{gathered} \tag{7}$$

The eq. (5) , (6) and (7) build up a system of diophantine equations that we call “ $S^{\text{TC}}$ systems” (see Ref. [8]). The solutions $\{m_r, v_r\}$ number of this systems varies with TC and quickly grows exponentially. No explicit solution formula of an $S^{\text{TC}}$ system is known to the author. By computer analysis one can retrieve the solutions for TC values of some tenths, in a reasonable machine time. For example with TC = 26, the number of solutions is computed as 565168. With TC = 5 and Rmax = 4, the $S^5$ system equations and solutions are (last solution as per Fig. 6):

$$8\,m_1 + 4\,m_2 + 2\,m_3 + m_4 = 16$$

$$m_1 + m_2 + m_3 + m_4 = 5$$

$$\{\{m_1 \to 0,\ m_2 \to 3,\ m_3 \to 2,\ m_4 \to 0,\ v_0 \to 1,\ v_1 \to 2,\ v_2 \to 1,\ v_3 \to 0,\ v_4 \to 0\},$$
$$\{m_1 \to 1,\ m_2 \to 0,\ m_3 \to 4,\ m_4 \to 0,\ v_0 \to 1,\ v_1 \to 1,\ v_2 \to 2,\ v_3 \to 0,\ v_4 \to 0\},$$
$$\{m_1 \to 1,\ m_2 \to 1,\ m_3 \to 1,\ m_4 \to 2,\ v_0 \to 1,\ v_1 \to 1,\ v_2 \to 1,\ v_3 \to 1,\ v_4 \to 0\}\}$$

Note that, given a set of solutions $\{m_r\}$ , the corresponding $\{v_r\}$ set is obtained by eq. (4). Finally it can be shown that, defining:

$$Q_r = m_r + v_r \tag{8}$$

the following relations holds:

$$\sum_{r=1}^{\text{Rmax}} v_r = \text{TC} - 2 \tag{9}$$

$$\sum_{r=1}^{\text{Rmax}} Q_r = 2\,(\text{TC} - 1) \tag{10}$$

Note that the $Q_r$ set can be viewed as the set of the ArbO “critical decisions” whose result can be a safe or “deadly” end.

# 4. The application of the Fermi statistics method

There are some formal analogies between the arbitrary oscillator model above described and the quantum models studied in the last century by E. Fermi to derive its statistics for ideal gas of molecules and electrons. In both cases we deal with each other indistinguishable objects, that can be arbitrarily allocated into available empty positions over certain predefined levels and imposing only one item per position. In both cases, any possible allocation must comply with some ”boundary” conditions. In the next rows we follow the approach of the great scientist even using the same notations and rationale. This Fermi method is well described in his book “ Molecules, Crystals and Quantum Statistics”, Ref. [9]. The rationale is as follows:

- we seek for the most probable solution of a $S^{\text{TC}}$ system
- to achieve this, we consider the number of ways to distribute some $m$ events over the available $Q$ cells at each step level
- the most probable solution will have the maximum number of the above said ways, say $\prod$
- we look for this maximum, looking at the maximum of $\text{Log}(\prod)$ -that is the same thing- using the method of Lagrange multipliers
- finally, we consider the boundary conditions to fix the unknown parameters

Considering our ArbO model, the above said $\prod$ value will be:

$$\prod = \binom{Q_1}{m_1}\binom{Q_2}{m_2}\ldots\binom{Q_r}{m_r}\ldots$$

where the $m_r$ and, accordingly, the $Q_r$, are solutions of a $S^{\text{TC}}$ system and the symbol $\binom{Q_r}{m_r}$ represents the binomial coefficient. If we pass to the natural logarithms Log() function and using the Stirling approximation formula, we will have:

$$\text{Log}(\prod) = \sum_{r=1}^{\text{Rmax}} (Q_r \,\text{Log}\; Q_r - m_r \,\text{Log}\, m_r - (Q_r - m_r)\,\text{Log}\,(Q_r - m_r))$$

we now consider the eq. (4), where we extract the $m_r$ term from the $\sum$ symbol, obtaining:

$$v_r = 2^r - m_r - \sum_{t=1}^{r-1} 2^{r-t}\, m_t \tag{11}$$

now, remembering the (8), we have:

$$Q_r = 2^r - \sum_{t=1}^{r-1} 2^{r-t}\, m_t \tag{12}$$

We see that, in the above eq., the $Q_r$ variable is independent from $m_r$, thing that is reasonable since $Q_r = 2\,v_{r-1}$. Now we can move on to the search of the maximum of the $\text{Log}(\prod)$ searching the maximum of the expression: $\text{Log}(\prod)$ -a C1 -b C2, where a, b are undetermined constant coefficients (according to Lagrange multipliers method), and C1, C2 are the first member of eq. (5) and (7), also constant in total value. We search then for the null condition of the derivative vs $m_r$ of the expression :

$$\text{Log}(\prod) - a \sum_{r=1}^{\text{Rmax}} m_r - b \sum_{r=1}^{\text{Rmax}} 2^{-r}\, m_r \tag{13}$$

After some mathematical steps, we find -as the null condition- the expression

$$m_r = \frac{2^r - \sum_{t=1}^{r-1} 2^{r-t}\, m_t}{1 + e^{a+2^{-r}\, b}} \tag{14}$$

Recalling the (12) , we have:

$$m_r = \frac{Q_r}{1 + e^{a+2^{-r}\, b}}$$

To find the values of a and b coefficients, we sum over $r$ and, imposing the boundary conditions (5), (9) and (10) , we obtain, after some algebra, the following expression leading to the eq. (1) presented in the first section.

$$m_r = \frac{2^r - \sum_{t=1}^{r-1} m_t\, 2^{r-t}}{2^{-r}\,(\mathrm{TC} - 2) + 1}$$

If we introduce the parameter rF = log(TC-2)/log(2), we obtain the following set of equations that represent - in recursive way - the most “probable” solution set of a diophantine $S^{\mathrm{TC}}$ system. We use here the bold character to distinguish this particular solution set to the other possible solutions.

$$\begin{aligned}
\mathrm{rF} &= \frac{\log(\mathrm{TC} - 2)}{\log(2)} \\
\boldsymbol{m_r} &= \frac{2^r - \sum_{t=1}^{r-1} \boldsymbol{m_t}\, 2^{r-t}}{2^{\mathrm{rF}-r} + 1} \\
\boldsymbol{v_r} &= 2^r - \boldsymbol{m_r} - \sum_{t=1}^{r-1} \boldsymbol{m_t}\, 2^{r-t} \\
\boldsymbol{Q_r} &= 2^r - \sum_{t=1}^{r-1} \boldsymbol{m_t}\, 2^{r-t}
\end{aligned} \tag{15}$$

# 5. Properties of the most probable solution

From the eq.s (15) we see that:

$$\boldsymbol{m_r} / \boldsymbol{Q_r} = \frac{1}{1 + 2^{\mathrm{rF}-r}} \tag{16}$$

The form of the (16) leads to a Cumulative Logistic Distribution shape. This is shown in the following Fig. 7 for TC=100 000, where the dashed line intercepts the curve at 0.5 level and the r axis at rF point, with rF=16.6096.

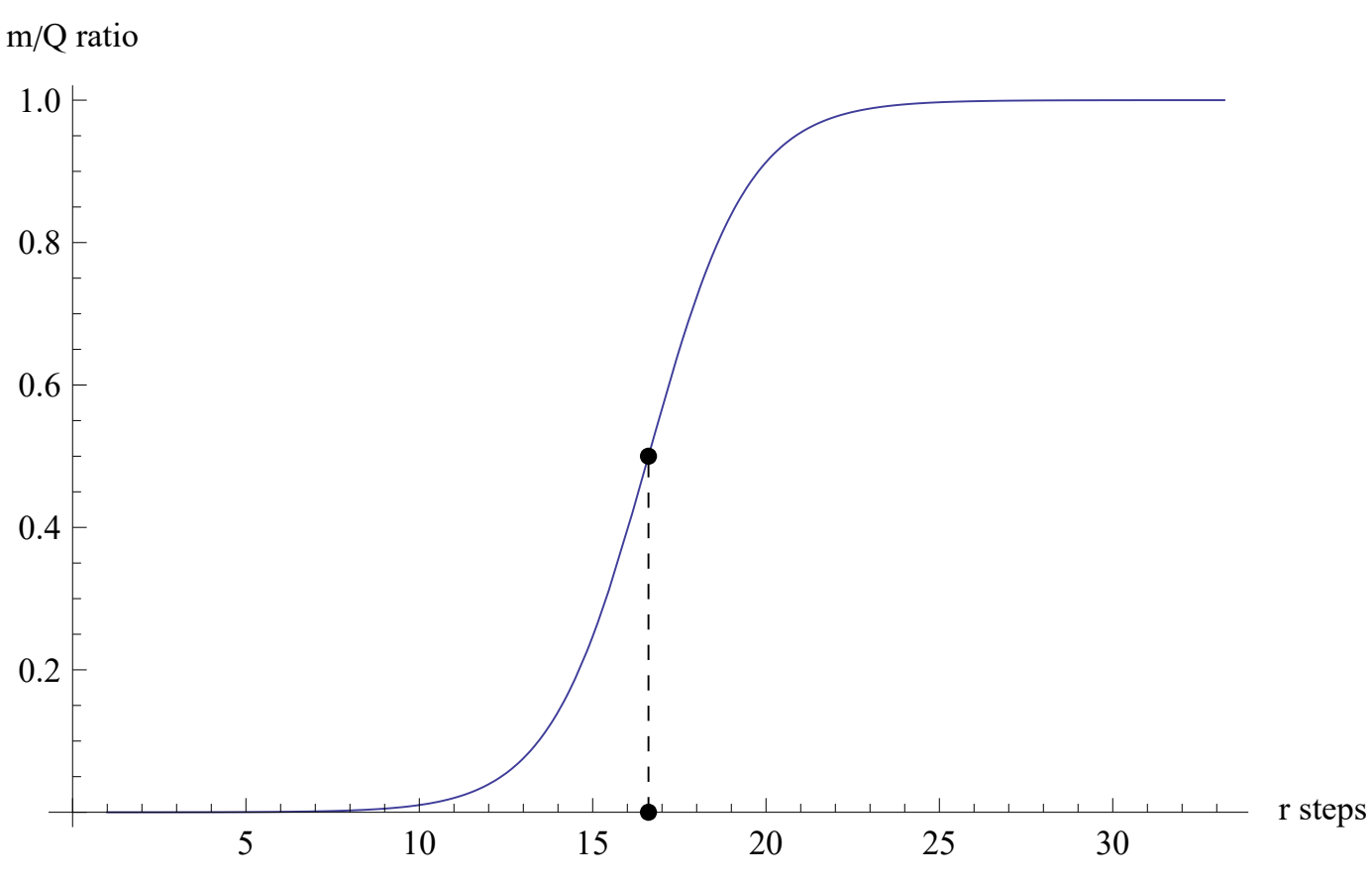

Fig.7- $(m_r / Q_r)$ Logistic shape for TC = 100000

The above curve says that , with 100 000 decisions cases, the most probable $\{\boldsymbol{m_r}\}$ solution of the $S^{\mathrm{TC}}$ system defined in eq. (7) will have the initial ***m*** values at zero, then, quite soon around rF, the ***m*** values become similar to the ***v*** values and -finally- all the ***m*** *(*EoL*)* values will saturate the available ***Q*** levels, meaning a life cycle end for the cellular automaton. It is seen that, even if Rmax=100 000-1, the most probable solution set will involve a number of non-zero variables limited to few tenths. This is indeed explained by the logarithmic dependence of rF vs TC (see eq. (15)). It is noteworthy to consider that will exist solutions with about 100 000 $m$ non zero variables, e.g. like the banal solution $\{m_1 = 1,\ m_2 = 1, \ldots, m_{99\,998} = 1,\ m_{99\,999} = 2\}$ } but these will have far less probability to arise. Let’s now consider the following Table 1, where, for the TC=100 000 case, the values of the **m, v, Q** (rounded to integer value) are presented together with the mortality figures for years 1974 and 2019 in Italy (ISTAT dx data and 2019 qx data), Ref. [7]. The calculation for the **m, v, Q** is done by computer, via iteration across the r steps , according to eq. (15).

| Years interval | r | m | v | Q | ISTAT - 2019 1000 x qx | ISTAT-2019 dx | ISTAT-1974 dx |
|---|---|---|---|---|---|---|---|
| Up to 4 years | 1 | 0 | 2 | 2 | 3.3453 | 335 | 2735 |
| 5-9 | 2 | 0 | 4 | 4 | 0.36563 | 36 | 180 |
| 10-14 | 3 | 0 | 8 | 8 | 0.43792 | 44 | 170 |
| 15-19 | 4 | 0 | 16 | 16 | 0.99425 | 99 | 334 |
| 20-24 | 5 | 0 | 32 | 32 | 1.429 | 142 | 385 |
| 25-29 | 6 | 0 | 64 | 64 | 1.63059 | 162 | 368 |
| 30-34 | 7 | 0 | 128 | 128 | 1.9993 | 198 | 505 |
| 35-39 | 8 | 1 | 255 | 255 | 2.86143 | 283 | 686 |
| 40-44 | 9 | 3 | 507 | 509 | 4.69355 | 463 | 1128 |
| 45-49 | 10 | 10 | 1003 | 1014 | 7.33727 | 721 | 1863 |
| 50-54 | 11 | 40 | 1966 | 2007 | 11.46247 | 1118 | 2799 |
| 55-59 | 12 | 155 | 3778 | 3933 | 18.46051 | 1780 | 4470 |
| 60-64 | 13 | 572 | 6984 | 7556 | 29.60728 | 2801 | 6265 |
| 65-69 | 14 | 1966 | 12002 | 13968 | 47.28093 | 4341 | 9089 |
| 70-74 | 15 | 5924 | 18079 | 24003 | 75.57279 | 6611 | 12542 |
| 75-79 | 16 | 14315 | 21843 | 36158 | 130.66618 | 10566 | 16581 |
| 80-84- | 17 | 24780 | 18906 | 43686 | 227.36915 | 15984 | 17535 |
| 85-89 | 18 | 27370 | 10441 | 37811 | 404.2388 | 21956 | 13839 |
| 90-94 | 19 | 17537 | 3345 | 20882 | 621.66989 | 20117 | 6806 |
| 95-99 | 20 | 6107 | 582 | 6690 | 798.55747 | 9776 | 1591 |
| 100-104 | 21 | 1112 | 53 | 1165 | 931.49213 | 2297 | 127 |
| 105-109 | 22 | 104 | 2 | 106 | 986.30673 | 167 | 2 |
| 110-114 | 23 | 5 | 0 | 5 | 998.26306 | 2 | 0 |
| 115-119 | 24 | 0 | 0 | 0 | 999.85919 | 0 | 0 |
| Total | | 100001 | 100000 | 200002 | | 99999 | 100000 |

Table 1. m, v, Q values with ISTAT 1974 & 2019 data for 100000 Total Cases

The Tab.1 data for **m, v, Q** are graphically shown in Fig. 8 here under, utilizing a suitable computer interpolation algorithm to generate continuous lines. The dashed line starts at rF and intercepts -as expected- the crossing of **m** and **v** curves. It can be demonstrated that the peaks of the three curves conserve their relative allocation around rF value and between themselves, for any rF value.

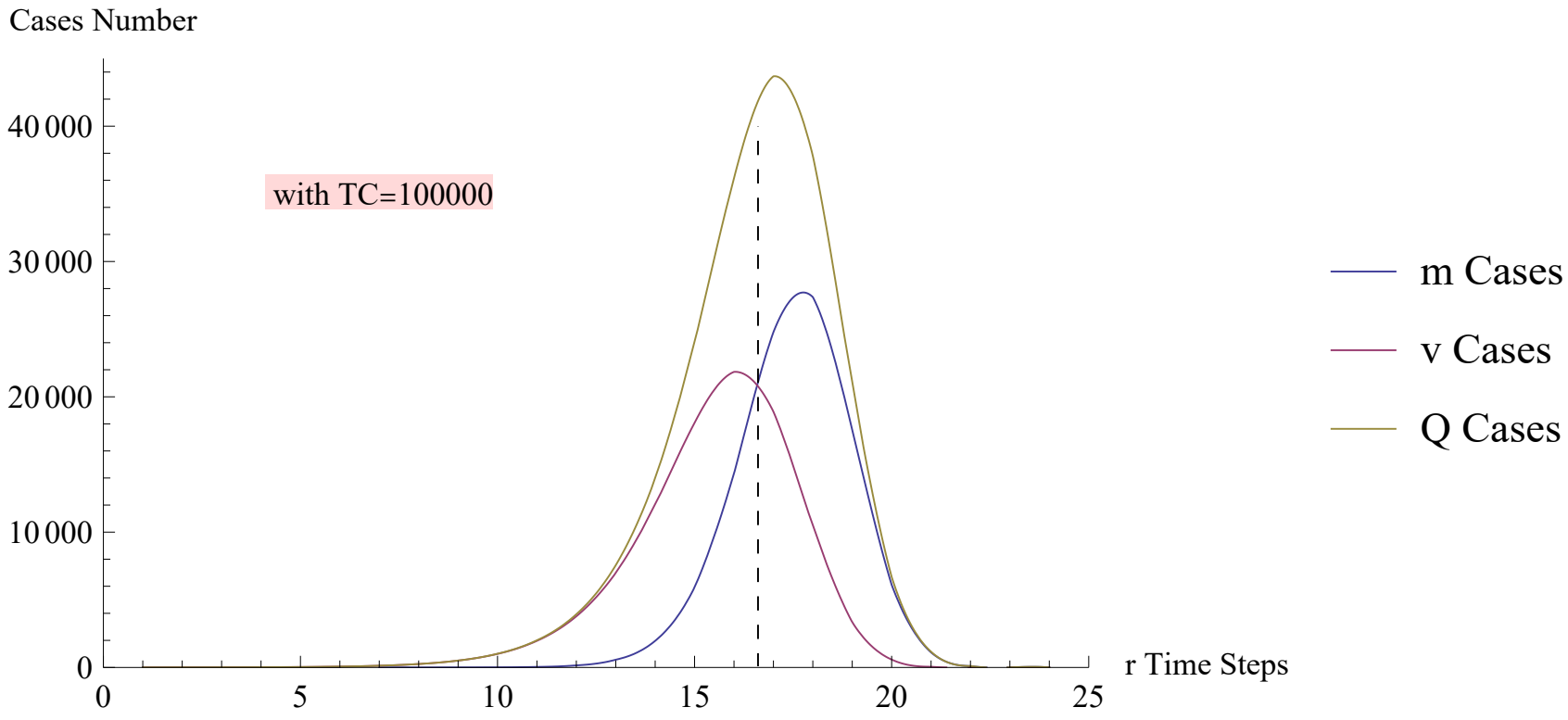

Fig. 8 Interpolated m, v, Q curves

# 6. Comparison with demographic data

From Table 1 and Fig. 9, we see that the **m** data have the same peak interval and trailing slope of the ISTAT dx 2019 data. It is also evident that the infant and mid-age mortality components are not present in the **m** model. The “improvement” in the life span between years from 1974 and 2019 is also remarkable.

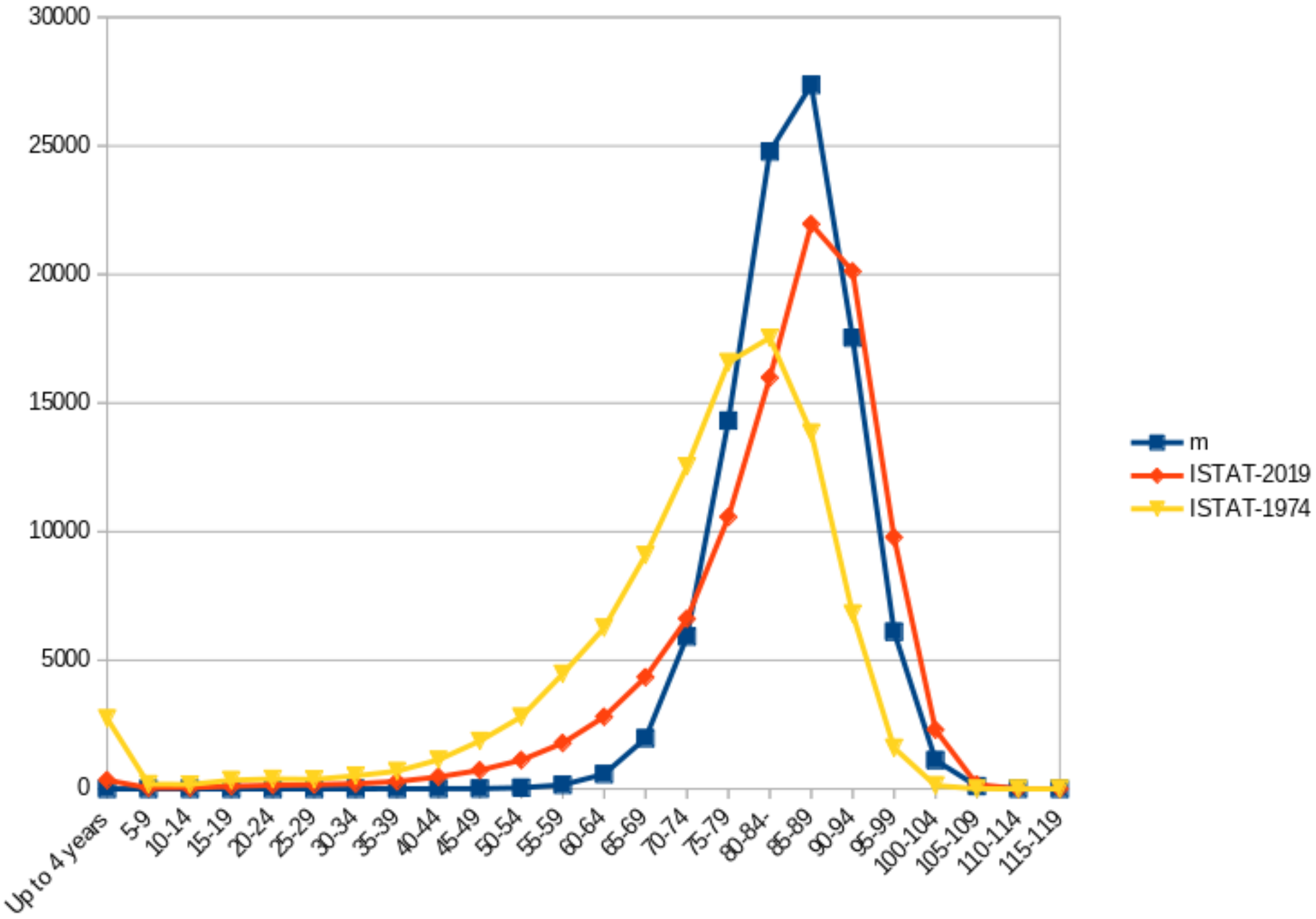

Fig. 9 Graphic comparison of m, ISTAT74, ISTAT19 curves vs five years intervals

To appreciate the meaning of the above comparisons, we must recall that the **m** curve in Tab.1 represents the most probable single solution of the $S^{100\,000}$ system. In other words - for a single ArbO subject- a plenty of EoL events can happen over the overall available Q events, accordingly with the possible "space paths". If we now consider a big number of samples of 100 000 ArbO objects, the various (most probable) alternatives in the life cycle paths will statistically materialize over the plotted **m** curve. Therefore this **m** curve becomes comparable with the standard mortality curves detected by the demographic survey relevant to the same normalized sample number of elements. This comparison is applicable if a particular size of age interval is chosen, namely the five years interval, corresponding to the unitary r step of the recursive model.

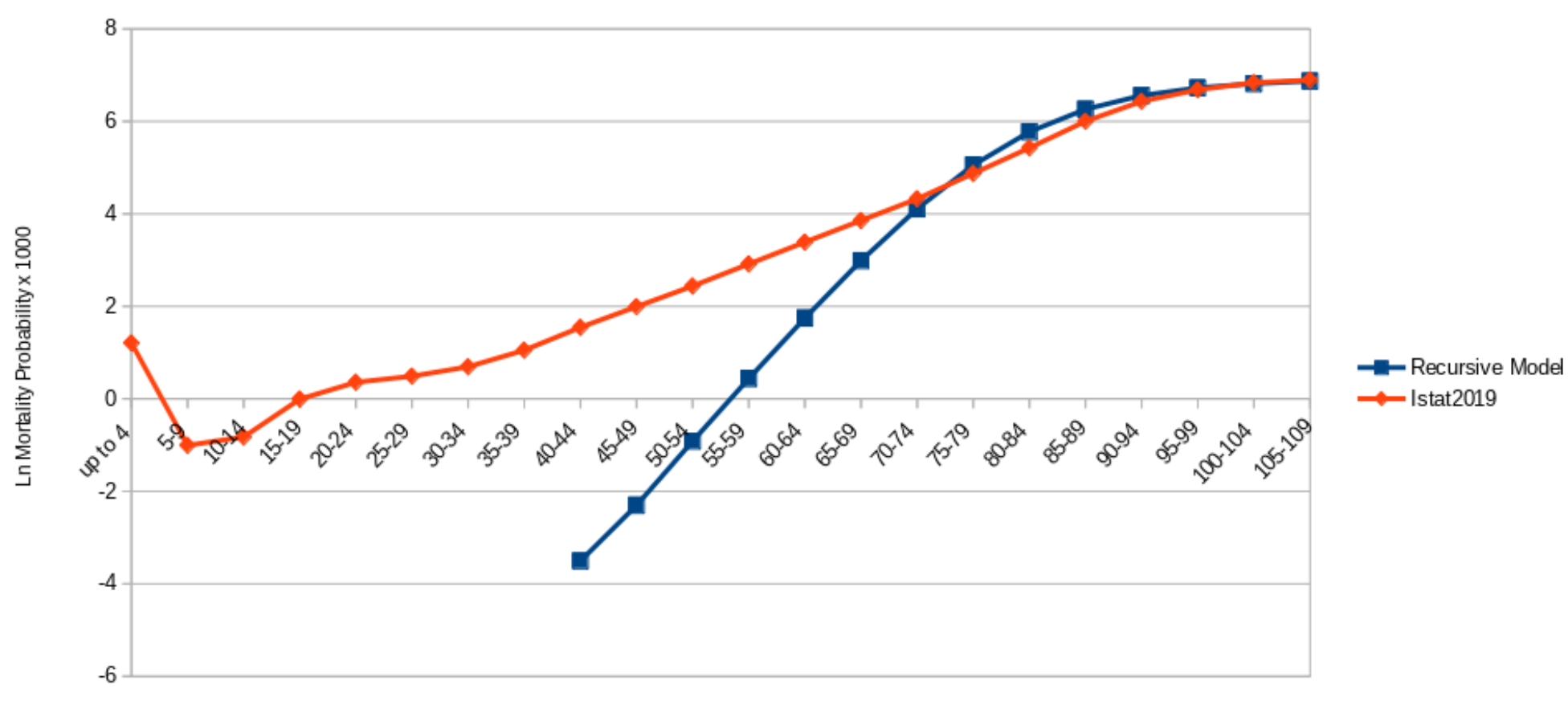

Fig. 10 Log mortality probability given by recursive model and ISTAT2019 data

From the Table 1, another Fig. 10 can be derived, comparing the mortality probability (x 1000 on log. scale ) derived from the ArbO recursive algorithm with the values measured with ISTAT2019. The mortality probability (qx) is defined here as the ratio between the number of deaths in a time interval and the surviving elements entering at the beginning of the interval and is usually defined as qx in demographic data, Ref. [3][7]. The Fig. 10 shows a linear shape of the log. curves in most part of the curve range (Ref. [10-11]). At initial and middle ages -however- a strong difference is evident (thing that is also evident on above Fig. 9 for mortality). Moreover, we see a common flattening (for both model and ISTAT data) of the curves slope at the oldest ages .

# 7. The continuous model equations

## 7.1 Introducing the continuous functions

The above (15) equations are finite difference equations, that require -for a numerical evaluation- a recursive computation. We want to transform these in explicit expression of the **m, v, Q** variables as functions of time and TC parameter. This can be useful for a better model handling and simulation. We look therefore for the differential equations possibly associated to the (15) equations and their solution. In doing this we introduce the ***m(t), v(t), Q(t)*** functions of a continuous time t variable such that:

$$\boldsymbol{m}_r = \boldsymbol{m}(r);\ \ \boldsymbol{v}_r = \boldsymbol{v}(r);\ \ \boldsymbol{Q}_r = \boldsymbol{Q}(r);$$

The above conditions imply that the continuous curves **m, v, Q** pass through the discrete points $(r,\boldsymbol{m}_r)$, $(r,\boldsymbol{v}_r)$, $(r,\boldsymbol{Q}_r)$ respectively, individuated by the recursive equations (15) (*). We keep the bold face of the function character meaning that we are again handling the particular (most probable) solutions set of (15) equations. Since the ArbO model is essentially a discrete binary model, the above defined continuous functions do not have a “physical” meaning (like e.g. number of choice cells, etc) but they can have a statistic meaning if we imagine a plenty of ArbO objects starting at the same time and evolving independently. In this case the **m, v, Q** curves can quantify the mean number of the corresponding random choices and events along a sliding time interval.

---

(*) The constraint assumed for the continuous curve implies a possible shift (about 2.5 years) to the right of the simulated curve with respect to the real curve. This is not critical for our analysis, as we are looking for structural effects (not influenced by an offset component) and not for an absolute age effect.

Considering the equation (2), we know that, in a continuous $t$ domain and with $\boldsymbol{m(t)} = 0$, the equation becomes equal to $\boldsymbol{v}' = k\boldsymbol{v}$, giving the solution $\boldsymbol{v(t)} = 2^t$ , with k = Log[2]. Extending the analysis including $\boldsymbol{m(t)} \neq 0$, we consider the differential equation:

$$\boldsymbol{v}' = k\,(\boldsymbol{v} - \boldsymbol{m}) - g\,\boldsymbol{m}'\,;\;\; g = \text{const.} > 0$$

This equation still comes to (2) if $\boldsymbol{m(t)} = 0$ or if $\boldsymbol{m(t)} = \boldsymbol{v(t)} = const.$ For the general case $\boldsymbol{m(t)} \geq 0$, we try some possible expressions of the g constant looking for a possible best fitting solution vs the points calculated with our recursive model. We find that a possible good fitting is reached with a value such that (*Assumption (iii)*):

$$g = (1 - k)$$

giving the differential equation:

$$\boldsymbol{v}' = k\,(\boldsymbol{v} - \boldsymbol{m}) - \boldsymbol{m}'\,(1 - k) \tag{17}$$

## 7.2 Solving the equations

Equation (17) involves two unknown time functions $\boldsymbol{v(t)}$ and $\boldsymbol{m(t)}$. Recalling relation (16), which has no recursive form, we can extend this relation to the continuous domain. We thus obtain a dependence between $\boldsymbol{v}$ and $\boldsymbol{m}$ such that:

$$\boldsymbol{v}(t) = 2^{\text{rF}-t}\,\boldsymbol{m}(t) \tag{18}$$

Using such a relation, we can substitute the $\boldsymbol{v}$ terms in the (17) diff. equation thus obtaining a diff. equation in only one unknown function $\boldsymbol{m(t)}$. To avoid excessive reader burden, we jump directly to the computed solutions of the combined (17) and (18) equations. For the solution computation we define a boundary condition such as, in consistency with the eq. (7):

$$\int_0^\infty \boldsymbol{m}(t)\, 2^{-t}\, dt = 1$$

In Ref. [8] more details will be available. These computations are performed with the aid of a well known commercial mathematical computer application. These solutions have the following form:

$$\begin{gathered}\boldsymbol{m}(t) = 2^{2t - \frac{(1+2g)\,\text{Log}\left(2^{\text{rF}} + 2^t g\right)}{g\,k}}\ \text{c1} \\ \boldsymbol{v}(t) = 2^{\text{rF}-t}\,\boldsymbol{m}(t) \\ \boldsymbol{Q}(t) = \boldsymbol{m}(t) + \boldsymbol{v}(t)\end{gathered} \tag{19}$$

The g, k terms are numbers while c1 and rF are constant depending only from the TC parameter .

$$\begin{gathered}g = (1 - k)\ ;\ k = \text{Log}(2) \\ \text{c1} = (1 + g)\left(2^{\text{rF}} + g\right)^{1+\frac{1}{g}} k \\ \text{rF} = \frac{\text{Log}(-2 + \text{TC})}{k}\end{gathered} \tag{20}$$

With these assumption the (19) and (20) equations can be used to characterize the particular ArbO model presented in the Sect. 4 and 5, giving **m, v, Q** as functions only of time and of the TC parameter.

## 7.3 The continuous qx(t) mortality probability definition

In the case of the discrete ArbO model we compute the $qx_r$ probability values with the same method of the standard demographic tables, Ref. [3]-[7]:

$$\mathbf{qx}_r = \boldsymbol{m}_r \Big/ \left( \text{TC} - \sum_{j=1}^{r} \boldsymbol{m}_j + \boldsymbol{m}_r \right)$$

that means to compare the current r interval mortality figure □$\boldsymbol{m}_r$ with the survived subjects entering at the beginning of the r interval, i.e. the denominator value in the above formula. In our case the $\boldsymbol{m}_r$ figures are given by the ArbO algorithm, while the standard demographic tables use the dx values sampled in the demographic survey. For the ISTAT data we use the qx x1000 values already available in the standard tables and use this data as a reference for our model, just like the dx data, also available in the demographic tables. Note again that our $\mathbf{qx}_r$ calculation is an independent process from demographic data, to which it can be compared assuming a same sample TC value and a suitable r scale such that x [years] = 5 r.

With these assumptions in mind and the 7.2 subsection equations, we can also define the **qx**(t) continuous function as:

$$\boldsymbol{qx}(t) = \frac{\boldsymbol{m}(t)}{\boldsymbol{N}(t)}$$

where $\boldsymbol{N}$(t) is the survival subjects number at instant t (*Assumption (iv)*) :

$$\boldsymbol{N}(t) = \text{TC} - \int_0^t \boldsymbol{m}(\tau)\, d\tau + \text{gg}\, \boldsymbol{m}(t) \tag{21}$$

This equation comes from both the above discrete definition of qx and the introduction of a "trimming" term with a gg ( ≠ g) constant . The trimming term is due to the need to compensate the numeric precision when passing from the Sum function to the Integral function. The gg value will result, in practical calculations, as gg = 0.5 . In Ref. [8] an explicit form of $\boldsymbol{N}$(t) is given. This form has been obtained solving the integral in (21) using the (19) formulas.

## 7.4 The simulated curves

With the above foundations, we show some graphics (see Fig.11 and subs.) with the defined ***m(t), v(t), Q(t), qx(t)*** functions over-imposed to points computed with the recursive algorithm based on (15) relations, when a TC value of 100 000 is assumed. The *t* values runs continuously along the *r* integers steps range. The dots coincide, except for minor deviations, with the $\boldsymbol{m}$(t) curve. We see that the interpolation shapes of the numerical data given in Tab. 1 with Fig. 8 are quite confirmed, and also that the area under the curves results to be close to the expected sizes (eq. (5) & (9)). The “dotting” of $\boldsymbol{Q}$ curve is omitted to avoid graphic burden to the Figures.

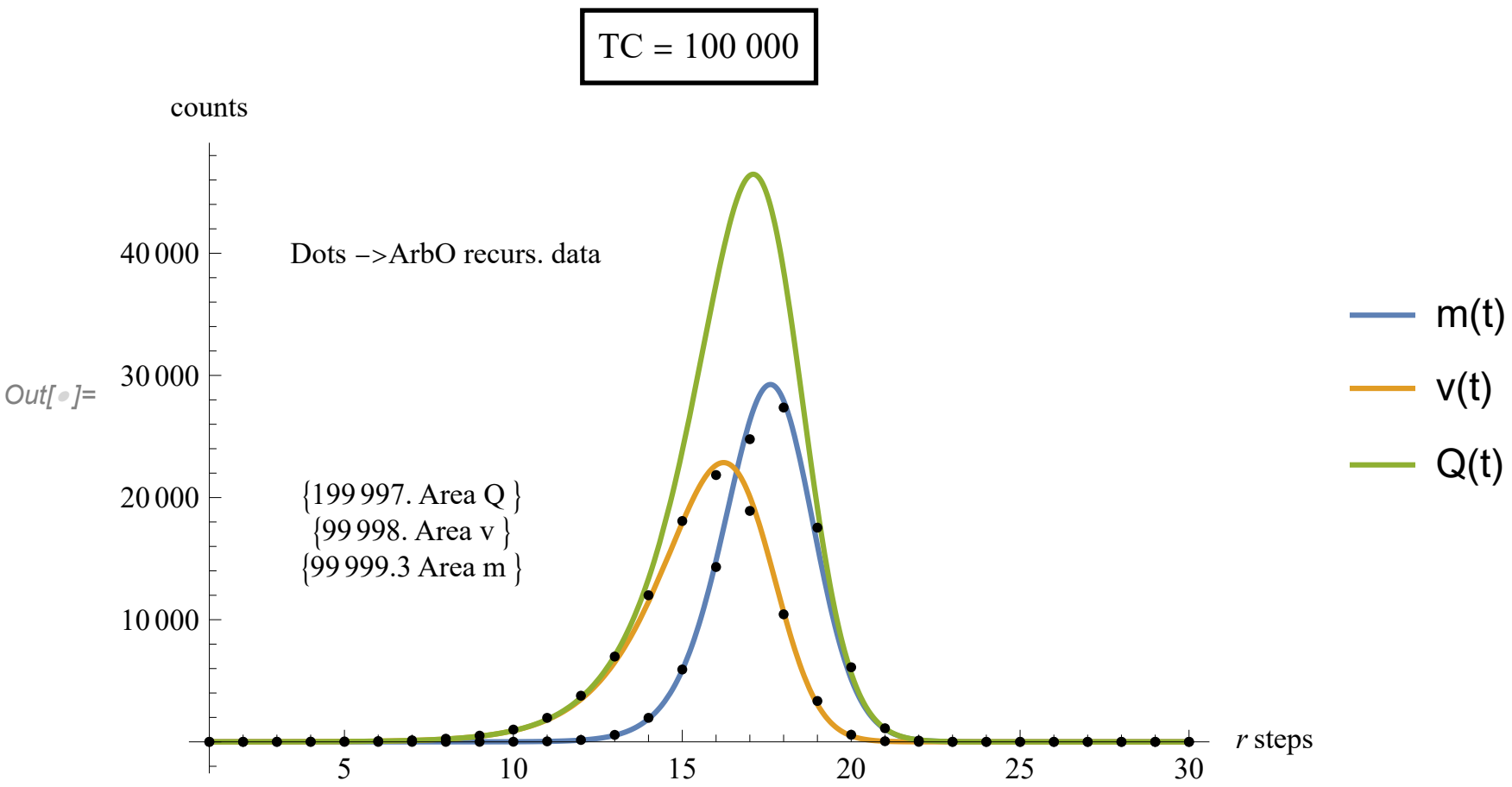

Fig.11 Comparison between the continuous curves m,v,Q and the recursive ArbO discrete data, TC=100 000

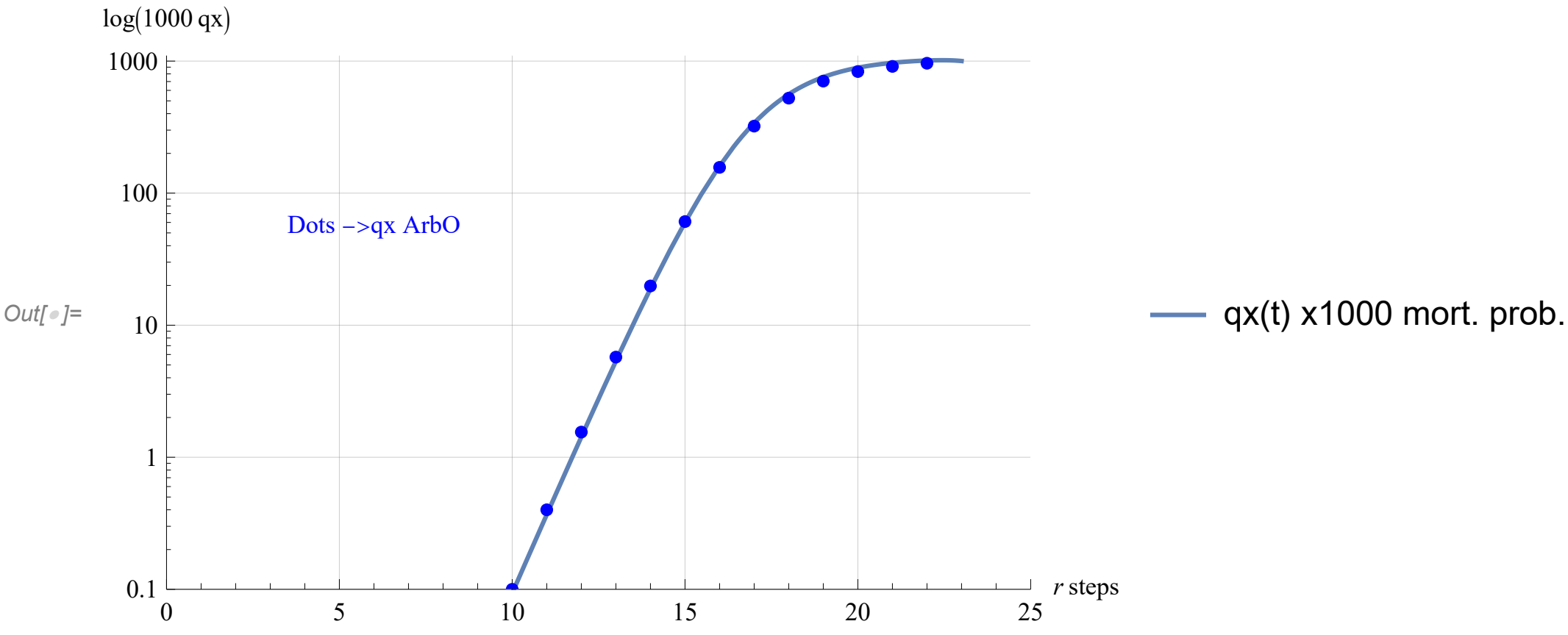

Fig.12 Comparison between the continuous qx curve and the ArbO discrete data, TC=100 000

We see also from Fig. 12 that the ***qx***(t) simulated curve fits well with the computed ArbO model qx dots (same blue dots on the Fig. 10). The shape of the theoretical ***qx***(t) confirms also a flattening at high ages.

## 7.5 The m, v, Q, qx curves features vs different TC parameter

Until now we considered a reference fixed value of TC = 100 000. It could be interesting to look at the effects on the curves of a variable TC . We found that the curves confirm the same shape and aspect, as per Fig. 13 and Fig. 14, with e.g. TC =10 000 and 1000 000 respectively. We also see that the curves slide along the x-axis due to their different rF value (located under the crossing of **m** and **v** curve) and that these have accordingly different areas as per the TC specific value. A similar analysis can be done for the theoretical ***qx***(t) magnitudes resulting in the Fig. 14. These show a convergence at advanced ages of the qx probabilities. The wide span between the curves is due to the three order of magnitudes considered for the TCs. All these figures show theoretical -continuous & discrete- data coming from the ArbO model.

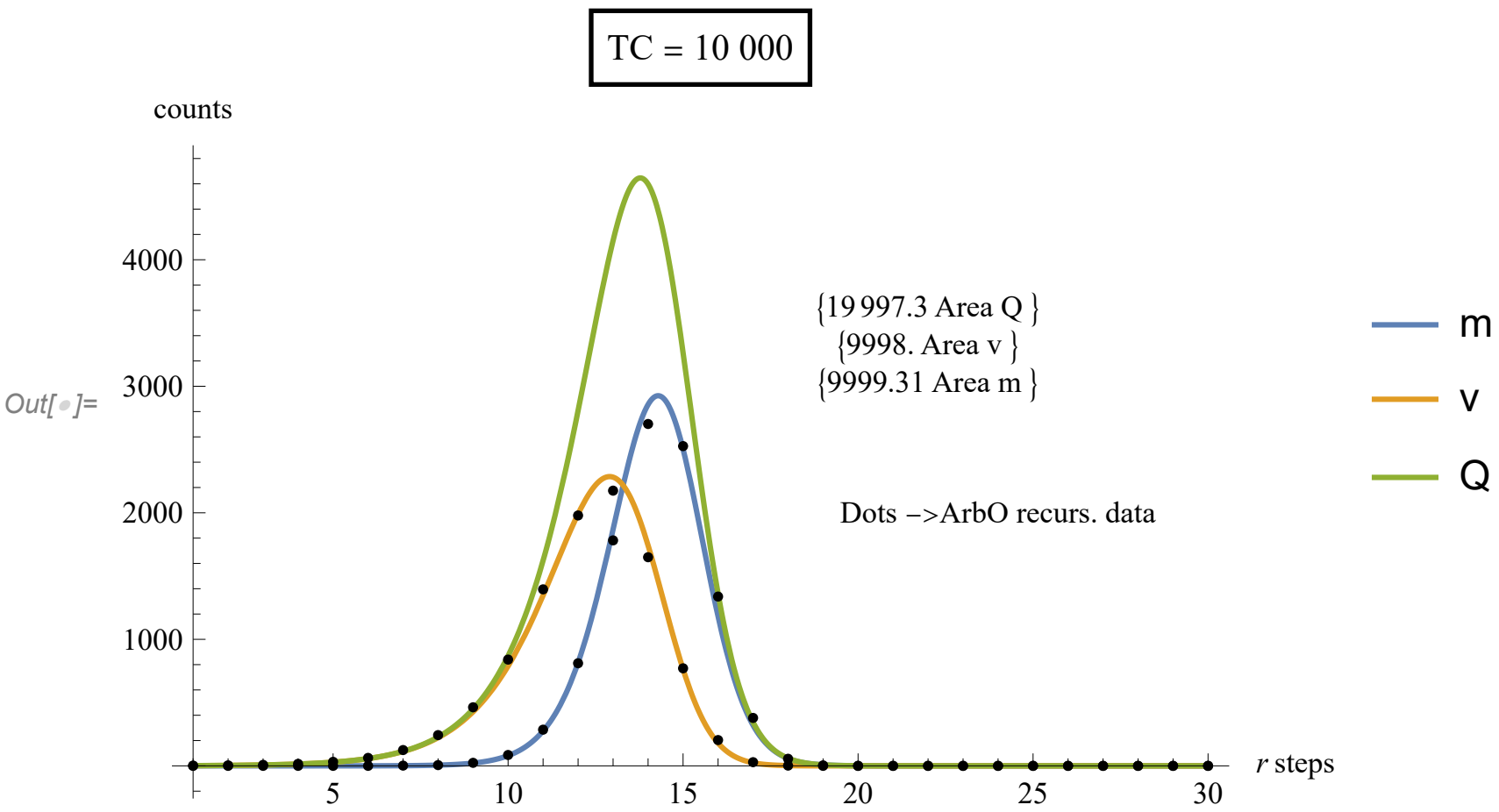

Fig. 13 Comparison between the continuous curves m,v,Q and the recursive ArbO discrete data, TC=10 000

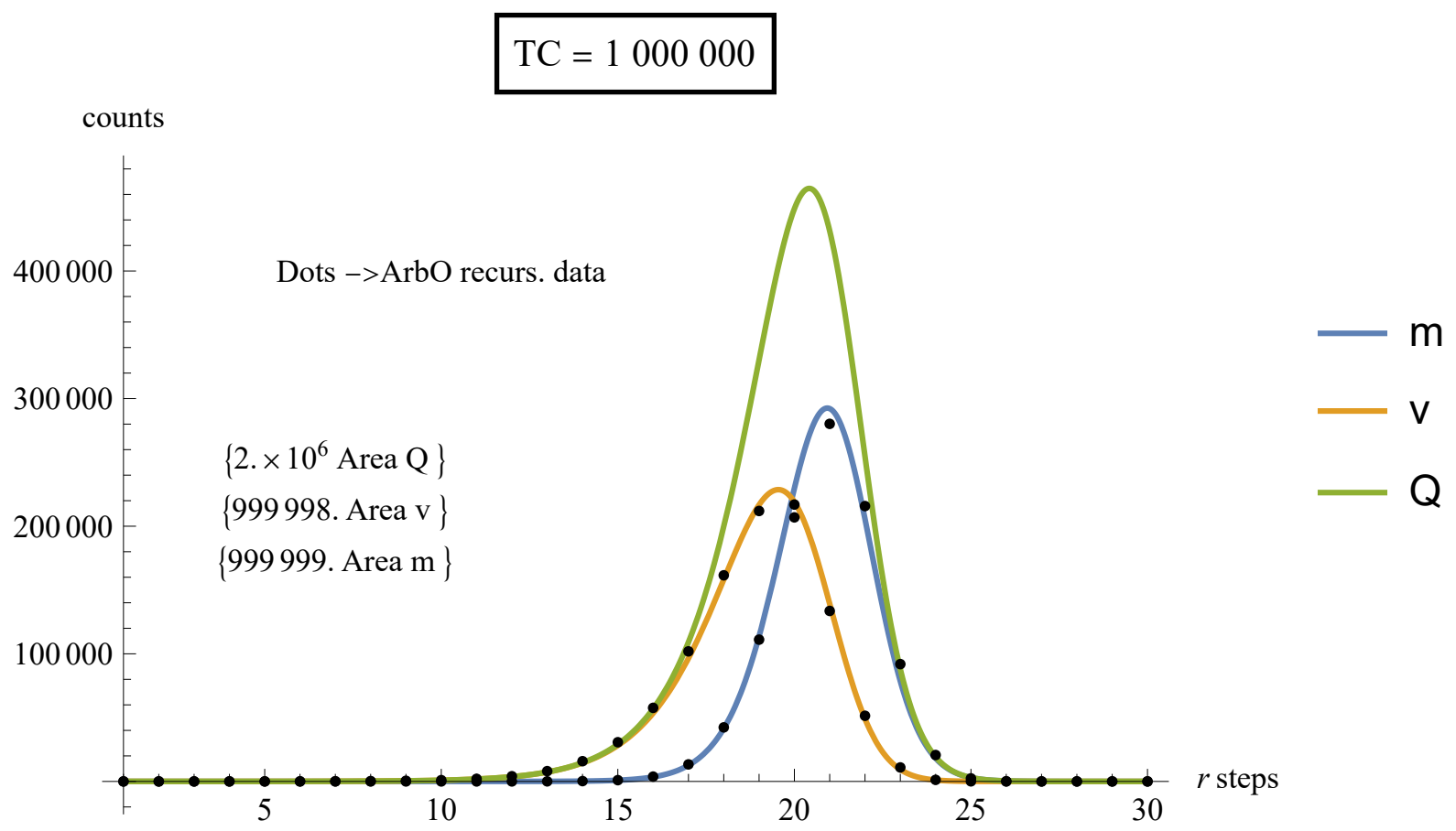

Fig.14 Comparison between the continuous curves m,v,Q and the recursive ArbO discrete data, TC=1 000 000

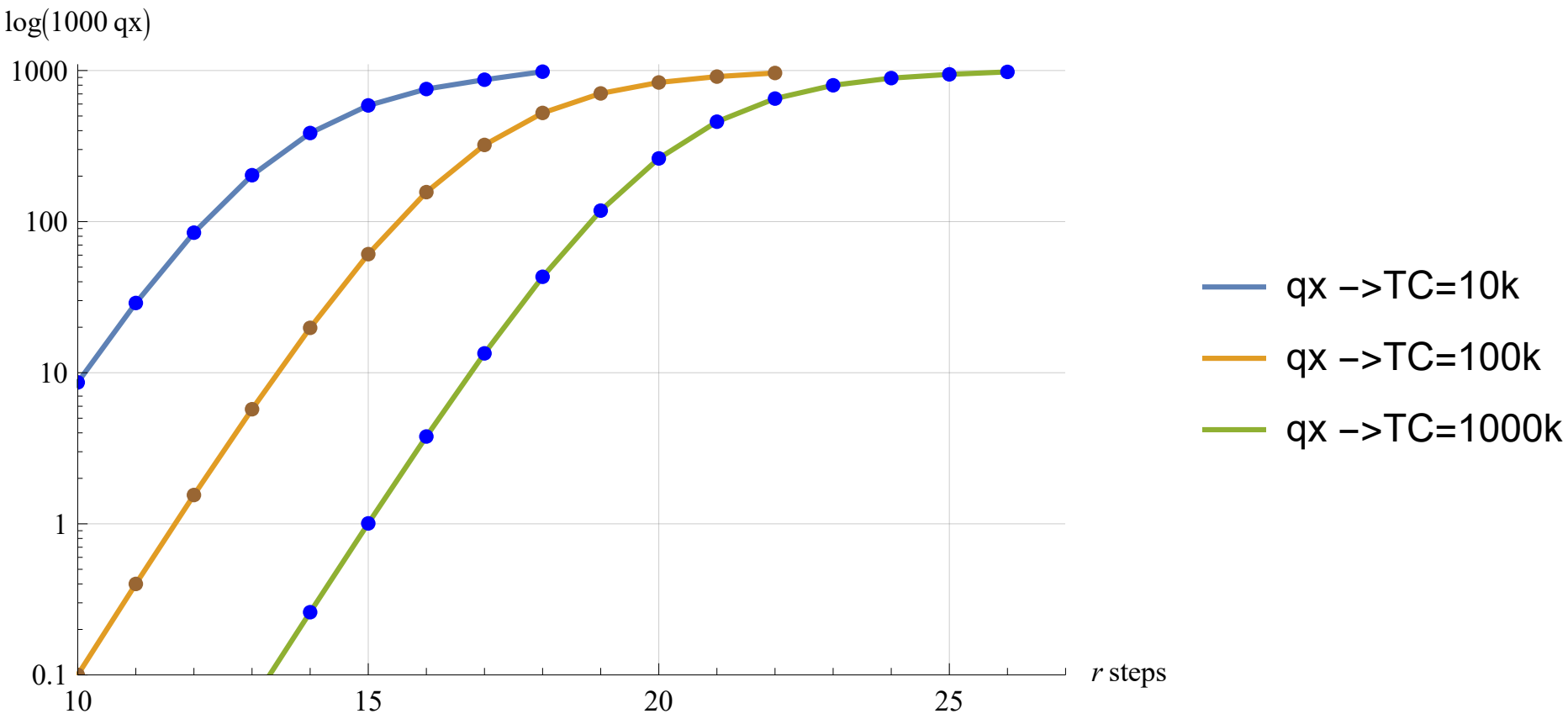

Fig.15 Theoretical qx(t) values computed with the ArbO model for different TC parameter values

It is also interesting to consider the question of how the "peaks" of these curves can change along different TC values. This is done calculating the curves null derivative position on the r step axis (giving a fractional real r value). The result is shown in Fig.16, where the positions of the curves peaks (presented in years units) are plotted vs TC range. It is seen that the relative peaks and rF level positions distances result constant. The figure shows also the logarithmic shape of the bundle since rF = log(TC-2)/log(2). The Fig.16 includes also a line called "lifespan" defining the position of the r value when the m curve reach the 10% (at the "older ages" right side) of the peak value . It happens that this "delta" value to the ***m*** peak is also constant resulting in about 2.77 r intervals (i.e. about 13.8 years from the max mortality peak position).

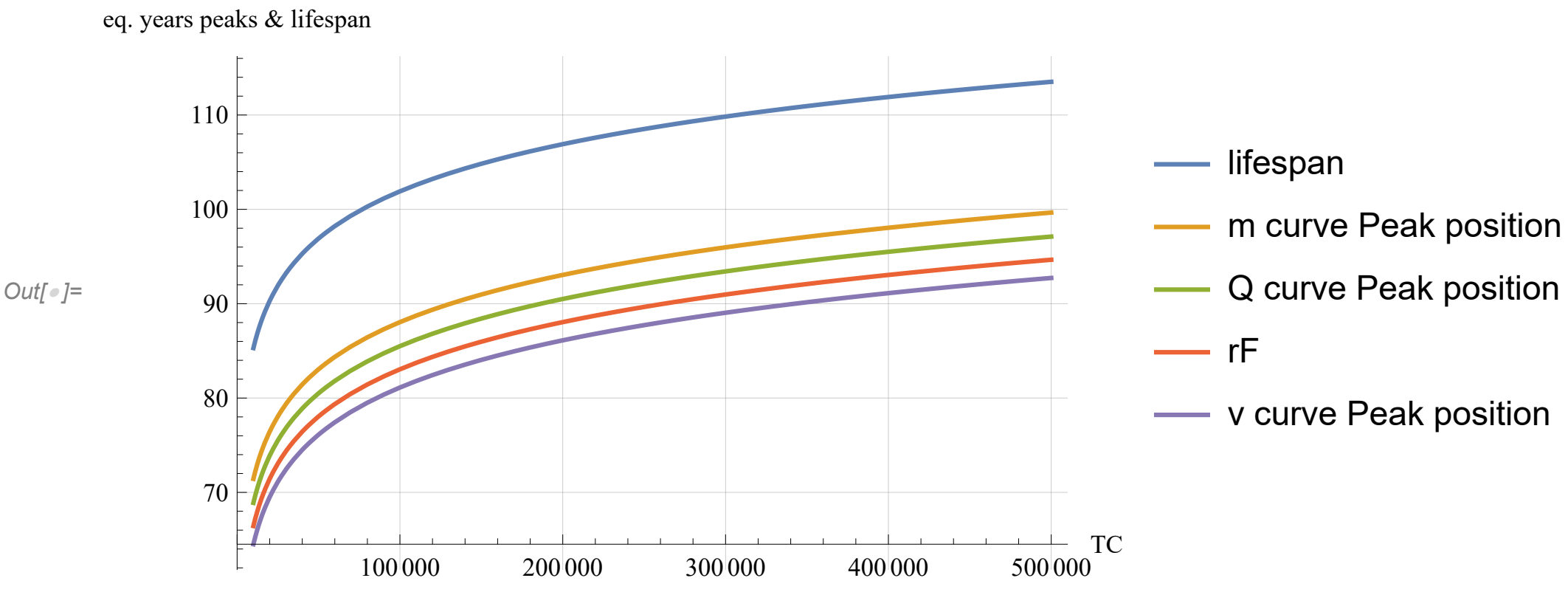

Fig.16 Peak positions for the continuous curves m,v,Q vs TC values

## 7.6 A possible interpretation of the real demographic mortality curves

With the above theoretical foundations, if we look to Tab. 1 and Fig. 9, we see that the demographic mortality tables show two different peaks for ISTAT1974 and ISTAT2019 data respectively. Now, considering the ISTAT2019 case, we assume that the actual data curve is composed of the mixture of two groups of subjects with different TC values coexisting in the demographic survey. Figures 17 and 18 show the total combined effect and the decomposition into individual curves to be summed to obtain the total effect. For this exercise we consider two TC values groups with different percentage weights on the total normalized area of 100 000 events. These $w_j$ weights can be defined as follows, where the $p_j$ numbers represent the share portions of the groups "labeled" as $TC_j$ that constitute the overall total T of subjects; these $w_j$ values therefore will "normalize" the total sample under analysis to 100 000 cases:

$$\sum_{j=1}^{n} p_j \, \mathrm{TC}_j = \mathrm{T}; \;\; w_j = (100\,000\, p_j)/\mathrm{T}; \;\; \sum_{j=1}^{n} w_j \, \mathrm{TC}_j = 100\,000$$

thus the "mixed" mortality profile will be given, in the case of two components, by:

$$\boldsymbol{mmix}(t) = w_1 \, \boldsymbol{m}(t, \; \mathrm{TC1}) + w_2 \, \boldsymbol{m}(t, \; \mathrm{TC2})$$

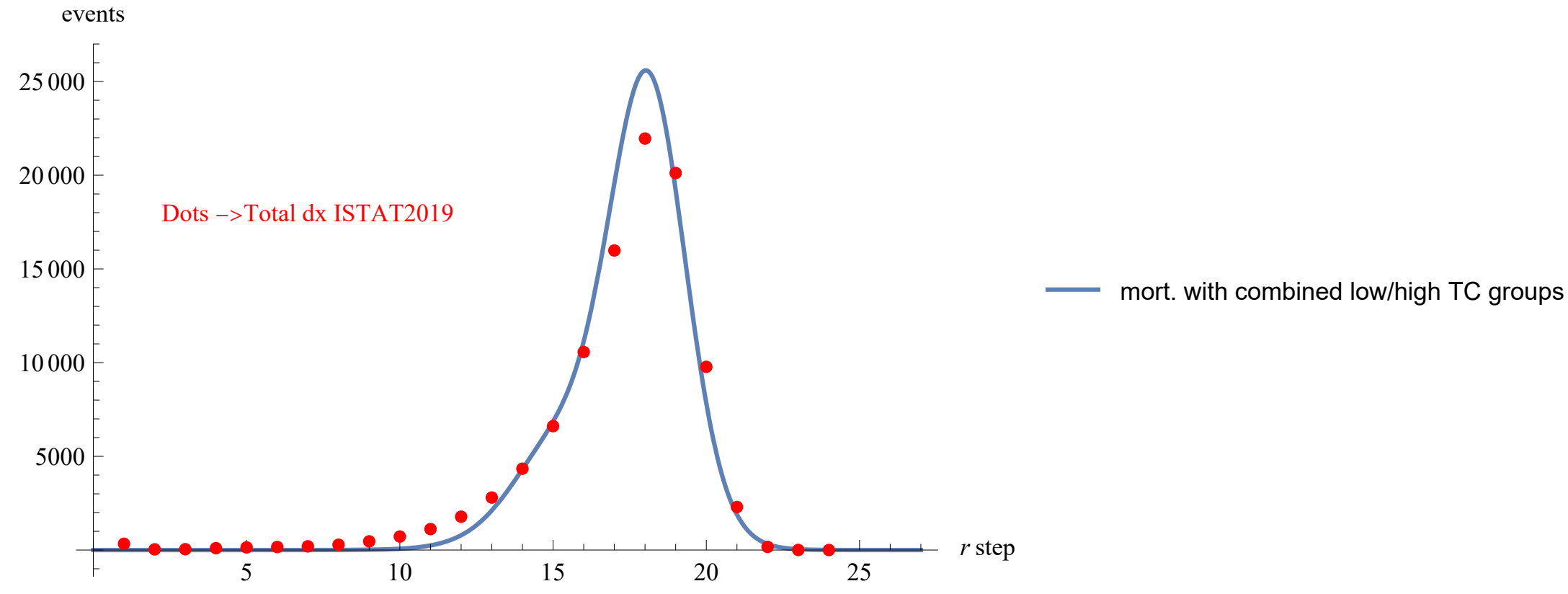

Fig.17 ISTAT2019 mortality data over a curve combined by sub-groups with different TC

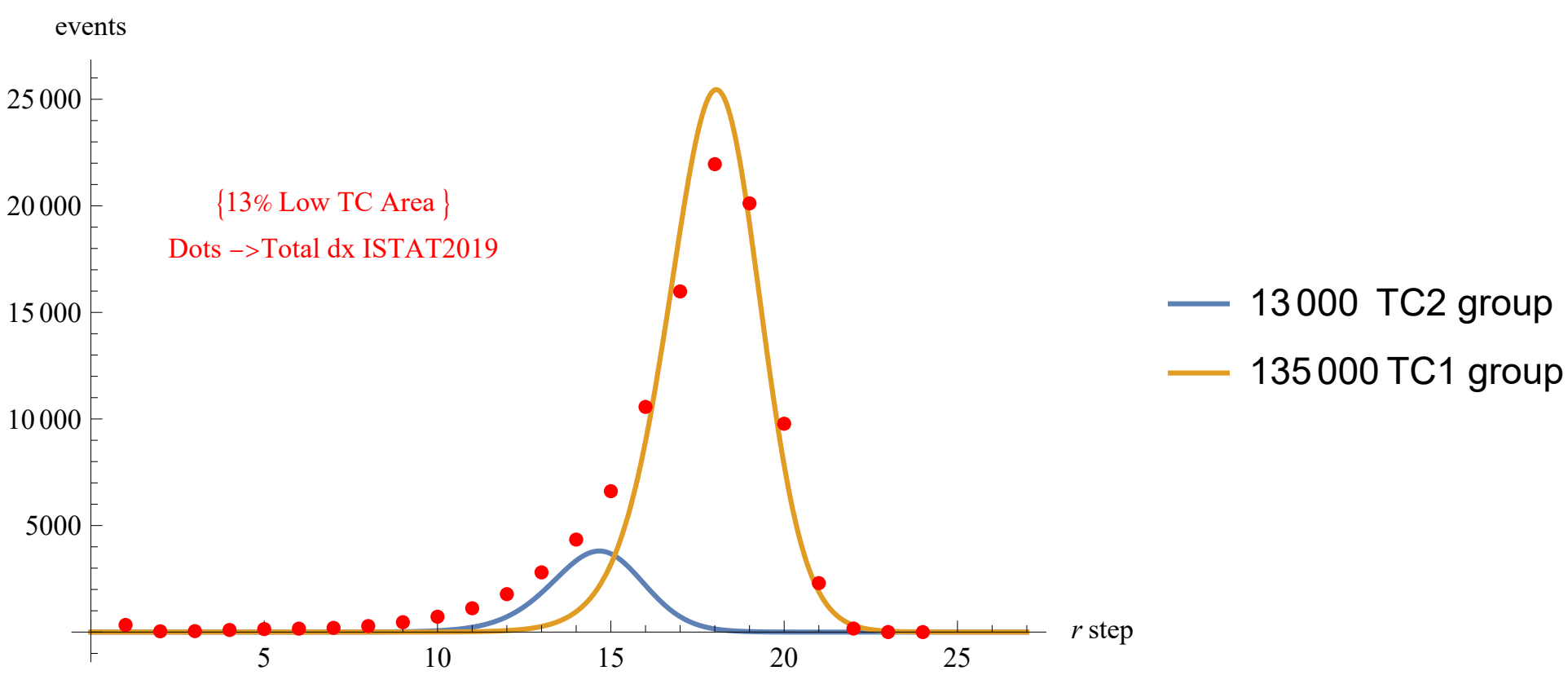

Fig.18 ISTAT2019 mortality data and the two sub-groups curves to be summed to give the combined total curve

The same approach is applied to the more elongated curve of ISTAT1974. With the aid of three combined curves with different TC values and groups weights we give the Fig. 19 and 20.

Fig.19 ISTAT1974 mortality data over a curve combined by sub-groups with different TC

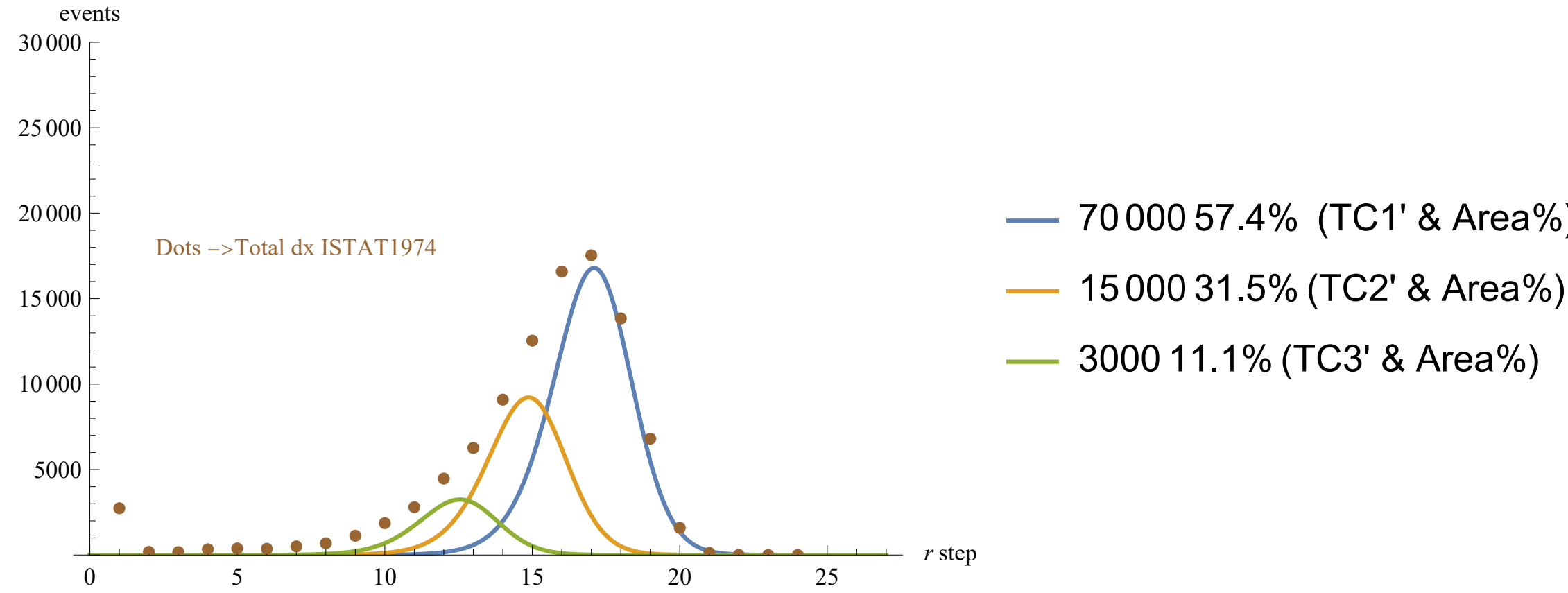

Fig.20 ISTAT1974 mortality data and the three component sub-groups curves to be summed to give the combined total curve

From the previous simulation tests we see that one possible explanation for the shape of the real mortality demographic curves is the overlap of mortality from different component groups. The same as above considerations can be extended to the qx probability as result of a mix of sub-groups components.

For this task we define the mixed qx curve when in presence of e.g. three major components, as:

$$\boldsymbol{qxmix}(t) = \frac{w_1\, \boldsymbol{m}(t,\ \mathrm{TC1}) + w_2\, \boldsymbol{m}(t,\ \mathrm{TC2}) + w_3\, \boldsymbol{m}(t,\ \mathrm{TC3})}{w_1\, \boldsymbol{N}(t,\ \mathrm{TC1}) + w_2\, \boldsymbol{N}(t,\ \mathrm{TC2}) + w_3\, \boldsymbol{N}(t,\ \mathrm{TC3})}$$

The formula is proposed for the three components case of Fig. 19 (ISTAT1974 data) using the same “weights” $w_j$ . The $\boldsymbol{m}(t,\mathrm{TC})$ and $\boldsymbol{N}(t,\mathrm{TC})$ functions are those obtained before with (19), (20), (21) applying the relevant TC parameter. The result is shown in Fig. 21 hereunder, where we found a good fitting of the curves from ages higher then about 55 years ( r = 11). In the Fig. 21 we also added the result of the same exercise for ISTAT2019 data with a ***qxmix***(t) function calculated as above but with only two components (as in Fig .17 situation).

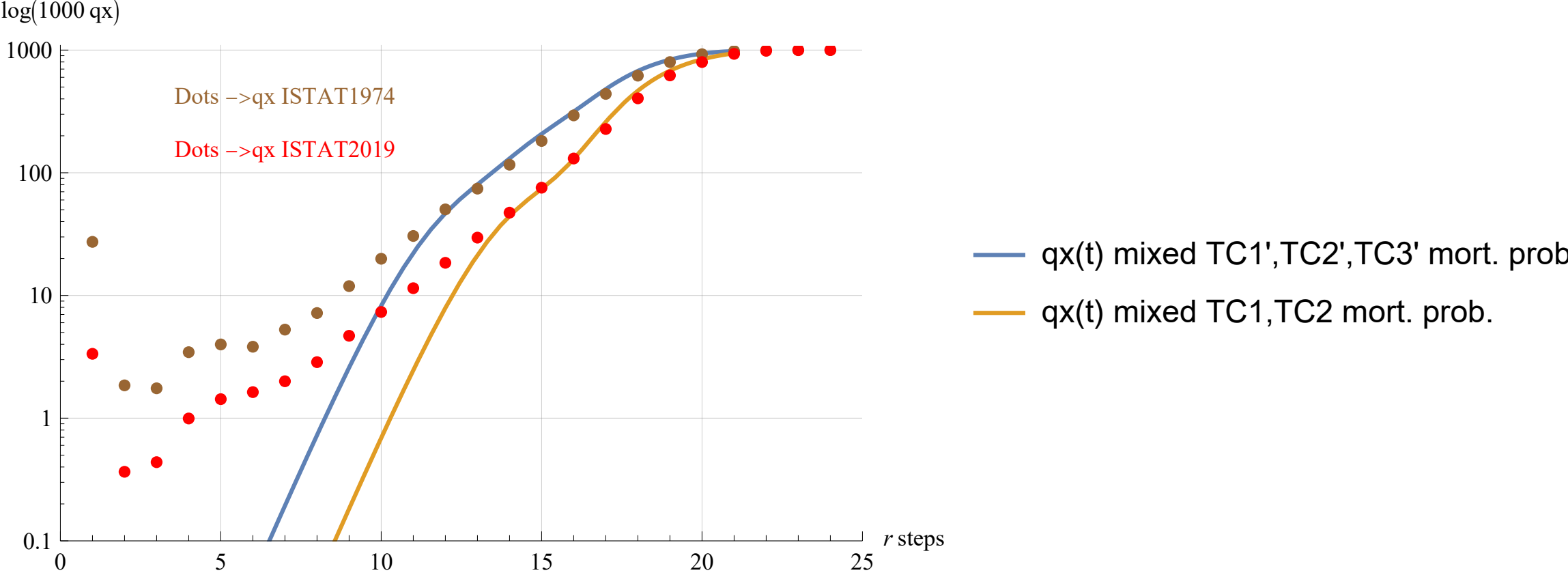

Fig.21 ISTAT1974 and 2019 mortality probability compared with qx(t) mixed curves coming from combining three groups or two groups respectively (same quotas as in Fig.20 and 18)

The above hypothesis of a multi-component structure of the curves derived from the dx and qx data in the demographic Life Tables was studied in more detail by the author in the work presented in Ref. [12].

# 8. Final comments and future analysis development

**General aspects of the model**

The present work considers and mathematically describes an abstract object (the Arbitrary Oscillator - ArbO) that generates events that evolve over time and may also include instances of "death." To find the most probable distribution of such events, the Fermi statistic is used, obtaining a recursive computational algorithm that provides the solution as dependent on a single parameter TC. By means of scaling, the mortality predictions generated by the algorithm can be compared with actual demographic mortality data. This comparison shows similarities of the theoretical curves generated by the algorithm with the shape of the real data.

**Specific aspects of the model**

- The mortality pattern that emerged from the study does not appear to have an absolute fixed limit to life span; this limit is set if the TC parameter is fixed but, as the TC parameter changes, the limit of maximum life expectancy shifts with logarithmic law (Fig. 16)
- Increasing the TC parameter lengthens the life span limit accordingly, but the critical phase narrows in percentage terms. This is due to the constant time distance (~14 years) expected between peak mortality and the end point of lifespan: for example, if the peak is at 80 years, one can hope to still live to 94 years (17% more); if the peak is at 100 years, one can hope to still live to 114 years (14% more).
- Flattening of the curve in the probability of death in old age is a matter of debate as to whether it is real or an effect of measurement methods and the sparsity and incompleteness of data (Ref. [11]). In the case of the proposed model, flattening is confirmed theoretically by the evolution of both recursive data and continuous simulation curves (Fig. 15).
- Actual mortality curves may be due to a mixture of components (sub-groups of subjects) with different TCs, these mixing of groups can explain the real curves of both mortality and mortality probability (Fig.17, 19, 21).

- If we look also to the evolution of the ISTAT curves, from past to today (Fig. 9), we see that the spread in the mortality peak is sharpening becoming closest to a ***m(t)*** curve. The model would seem indeed to predict that in more advanced social situations the possible different TCs groups tend to approach a single final TC value, while more backward social situations may be explained by a broader distribution of TCs groups.

**Open points and possible future developments**

When we associate an abstract object such as ArbO with real living subjects we can assume direct analogies with e.g. mortality events but less clear, in the analogy, is the meaning of the TC parameter and also of the **v** and **Q** variables. For the latter two variables we can give interpretations such as : number of “critical” decisions/events in a time interval (**Q** variable) and number of “safe-ending” decisions/events in the same interval (**v** variable).
For the TC parameter (“Total Cases”), there is no clear reference to the reality addressed in the analogy. In the ArbO model, the TC parameter limits the maximum number of paths in the “path space.” Limit reached which, all possible events must lead to the end of the life cycle. In sociological terms, our TC could represent a generic social target parameter such as ‘longevity’, whose improvement corresponds to a similar improvement in life expectancy.
An example of such ‘social’ parameter can be the ‘Intrinsic Capacity’(IC) , a concept introduced by the World Health Organization, in the context of the healthy ageing target. The IC parameter refers to an individual’s overall functional reserve, that is, the set of his physical and mental capacities that determine his well-being and autonomy as he ages.
A further possible extension of our model could be considered in the case of growth by splitting of biological or physical objects. In this case the model can be adapted to that of a system growing by splitting of components in the presence of a random risk of death and an absolute limit on the number of elements to be met (TC). On the latter aspect, i.e., population growth and the patterns that can govern it, references such as Ref.[12-15] can be considered and studied in correlation with future developments of our model.

# 9. Developments since the first version of this work

This article was prepared in May 2022 and has had several minor revisions at later times. Starting from the ArbO model presented here, the author subsequently developed a conjecture about the trend in demographic mortality at older ages. This conjecture was originally detailed in Ref. [16] and also led to a further study of the components of demographic mortality (developing an insight discussed in Section 7.6 of this paper). Finally, in a further paper, the author hypothesised the presence of absolute limits on mortality curves in demographic life tables. These last two papers are cited in Ref. [17-18] respectively.

## 9.1 Correlation with a recent study using a multi-omics approach

In a paper published in August 2024 in the journal Nature Aging (Ref. [19]), researchers highlighted a non-linear disease risk that accumulates with the biological ageing process. *"The analysis revealed consistent nonlinear patterns in molecular markers of aging, with substantial dysregulation occurring at two major periods occurring at approximately 44 years and 60 years of chronological age"* (cit. from [19]). This study involved a sample of US persons aged 75 years and under. In our studies (see above subsec. 7.6 and Ref. [17]), we showed the possible existence of discrete components of demographic mortality. These two approaches - conducted in completely different ways - show a correlation of results.

# References

*Author Note on References:*

*The literature on demographic mortality is enormous and covers a broad spectrum of scientific and social aspects. It is beyond the scope and capacity of this article to attempt to review and make reference with these studies. We will therefore only refer to works strictly instrumental to the contents of this work, leaving the correlation with other studies on the topic to the interest and experience of the reader. This choice can also be justified - in the author's opinion - by the fact that the approach illustrated here is original and not directly related to other works and research models.*